\documentclass[%
 reprint,
superscriptaddress,
 amsmath,amssymb,
 aps,
prb,
]{revtex4-2}

\usepackage{graphicx}
\usepackage{dcolumn}
\usepackage{bm}
\usepackage{orcidlink}
\usepackage{siunitx}
\usepackage{hyperref}

\begin{document}


\title{Direct imprinting of arbitrary spin textures using programmable structured light \\ in a semiconductor two-dimensional electron gas}

\author{Keito Kikuchi\,\orcidlink{0009-0006-5865-9416}}
    \affiliation{%
     Department of Materials Science, Tohoku University, Sendai 980-8579, Japan
    }%
\author{Jun Ishihara\,\orcidlink{0000-0002-9574-1332}}%
    \email{j.ishihara@tohoku.ac.jp}
    \affiliation{%
     Department of Materials Science, Tohoku University, Sendai 980-8579, Japan
    }%
\author{Miari Hiyama}%
    \affiliation{%
     Department of Materials Science, Tohoku University, Sendai 980-8579, Japan
    }%
\author{Sota Yamamoto}%
    \affiliation{%
     Department of Materials Science, Tohoku University, Sendai 980-8579, Japan
    }%
\author{Yuzo Ohno\,\orcidlink{0000-0002-7092-9793}}%
    \affiliation{%
     Graduate School of Pure and Applied Sciences, University of Tsukuba, Tsukuba 305-8573, Japan
    }%
\author{Takachika Mori}%
    \affiliation{%
     Department of Applied Physics, Tokyo University of Science, Tokyo 125-8585, Japan
    }%
\author{Kensuke Miyajima\,\orcidlink{0000-0002-7586-1409}}%
    \affiliation{%
     Department of Applied Physics, Tokyo University of Science, Tokyo 125-8585, Japan
    }%
\author{Makoto Kohda\,\orcidlink{0000-0002-1995-1282}}%
    \email{makoto.koda.c5@tohoku.ac.jp}
    \affiliation{%
     Department of Materials Science, Tohoku University, Sendai 980-8579, Japan
    }%
    \affiliation{%
     Center for Science and Innovation in Spintronics, Tohoku University, Sendai 980-8577, Japan
    }%
    \affiliation{%
     Division for the Establishment of Frontier Science, Tohoku University, Sendai 980-8577, Japan
    }%
    \affiliation{%
     Quantum Materials and Applications Research Center, National Institutes for Quantum Science and Technology, Gunma 370-1292, Japan
    }%

\date{\today}

\begin{abstract}
Precise control of spatial spin structures, such as spin helices, is critical for advancing spintronic devices, particularly in non-volatile, low-power information storage and processing. Conventional techniques, including transient spin grating spectroscopy and spatial- and time-resolved Kerr rotation microscopy, are limited by fixed optical grating periods and uniform light polarization, respectively, which constrain the flexibility of spin helix generation. Here, we introduce a novel approach utilizing structured light to directly imprint spatial spin structures in a GaAs/AlGaAs quantum well. This method allows for the precise control over the wave number and configuration of the spin helices, overcoming the limitations of previous techniques. Experiments conducted using pump-probe Kerr rotation microscopy combined with a programmable spatial light modulator revealed the efficient and tunable generation of spin helices. This approach is broadly applicable not only to semiconductors but also to magnetic thin films and 2D materials.
\end{abstract}

\maketitle


\section{INTRODUCTION}
The precise control of spin states through spin-orbit interactions is crucial in semiconductors. A persistent spin helix (PSH) state \cite{Schliemann2003-kj, Bernevig2006-lo,Koralek2009-mo, Kohda2012-mj, Walser2012-tg, Sasaki2014-ev, Kohda2017-lt} has garnered significant attention for its potential to generate spatial spin structures without external magnetic fields.The PSH state arises from the balance between Rashba \cite{Rashba1960-tf,Bychkov1984-iv} and Dresselhaus spin-orbit interactions \cite{Dresselhaus1955-sy} in III-V compound semiconductors, such as a GaAs/AlGaAs quantum well (QW). In this state, the spin precesses coherently over time, forming a helical spin structure that can be sustained over long distances. This spatial spin structure offers a stable, non-volatile medium for information storage and transport, making it a potential candidate for future spintronic devices \cite{Chuang2015-be, Kohda2017-lt, Dettwiler2017-ud, Kohda2023-jx}.

Historically, transient spin grating (TSG) spectroscopy \cite{Weber2007-py, Koralek2009-mo, Yang2011-rj, Yang2012-er, Yang2012-ue} was one of the first techniques used to generate spin helices in semiconductors. While TSG effectively produces a spin helix, its period is fixed by the pre-prepared optical grating itself, limiting the flexibility of this method in controlling the spatial period of the spin helix. Following TSG, spatial- and time-resolved Kerr rotation (STRKR) microscopy \cite{Walser2012-tg, Ishihara2013-eb, Salis2014-nf, Ishihara2014-xm, Altmann2014-wd, Chen2014-nt,  Altmann2016-tj, Kunihashi2016-rg, Passmann2018-hp, Anghel2020-wz, Anghel2021-ki, Ishihara2022-wg} was introduced as a more advanced technique, capable of studying spin dynamics with a high spatial and temporal resolution. However, STRKR typically relies on excitation with spatially uniform light polarization, which limits the speed at which spatial spin structures are created. This inherent limitation necessitates the development of more efficient methods for generating these spatial spin structures \cite{Ishihara2023-ep}.

In this study, we propose a novel method that extends the capabilities of STRKR by utilizing a spatial light modulator (SLM) to create structured light. By irradiating this structured light onto a GaAs/AlGaAs QW, a material system known for its excellent optical properties and for the one-to-one correspondence between polarization and spin polarization, we will demonstrate a method for selectively exciting the spatial spin structure corresponding to the structured light. This approach allows for a more efficient and flexible generation of spatial spin structures, offering greater control over the periodicity and spatial configuration. 

This method can be applied to systems with strong coupling between light polarization and spin polarization, so it is expected to be applicable to other material systems, not only to other III-V compound semiconductors \cite{Passmann2018-hp, Kawaguchi2019-kf, Arikawa2022-ep}, but also organic-inorganic halide perovskite semiconductors \cite{McIver2011-yd, Giovanni2015-vt, Boschini2015-mf, Huang2017-jw, Odenthal2017-me, Yumoto2022-ia, Anghel2023-re}, transition metal dichalcogenides \cite{Zhu2014-ff, Yang2015-ob, Yang2015-te, Plechinger2016-yw, Huang2017-zm, McCormick2017-tk}, 2D magnetic semiconductor \cite{Huang2017-tn, Huang2018-oe, Cenker2021-qu, Cheng2021-vu, Zhang2022-na}, magnetic thin films \cite{Lambert2014-xr, El-Hadri2016-tj, Takahashi2016-gq, John2017-uh, Kichin2019-kr}, and ferromagnetic/nonmagnetic interfaces \cite{Choi2020-xv, Iihama2021-gv}. Beyond stripe-like structures such as spin helices, the SLM could also be employed to transfer various forms of light information into solid-state systems, such as vector beams and holographic imaging, by spatially modulating the phase, polarization, and intensity of the light through spin.

\section{EXPERIMENT}
In a GaAs/AlGaAs QW, the optical selection rule \cite{Cameron1996-xq} allows for circularly polarized light (R-polarization and L-polarization) to transfer an angular momentum of ±1. This means that the light polarization corresponds directly to the spin polarization. In this study, we aim to generate a spin helix by spatially modulating the light polarization of the pump beam using a SLM in conjunction with the conventional STRKR microscopy technique. First, we investigated the extent to which the SLM can control light polarization and the spatial precision of this control. Based on these experimental results, we then assessed the feasibility of generating a spin helix.

\subsection{Sample and optical measurement system}
\begin{figure*}[htbp]\label{fig:fig1}
\centering
\includegraphics[width=0.7\linewidth]{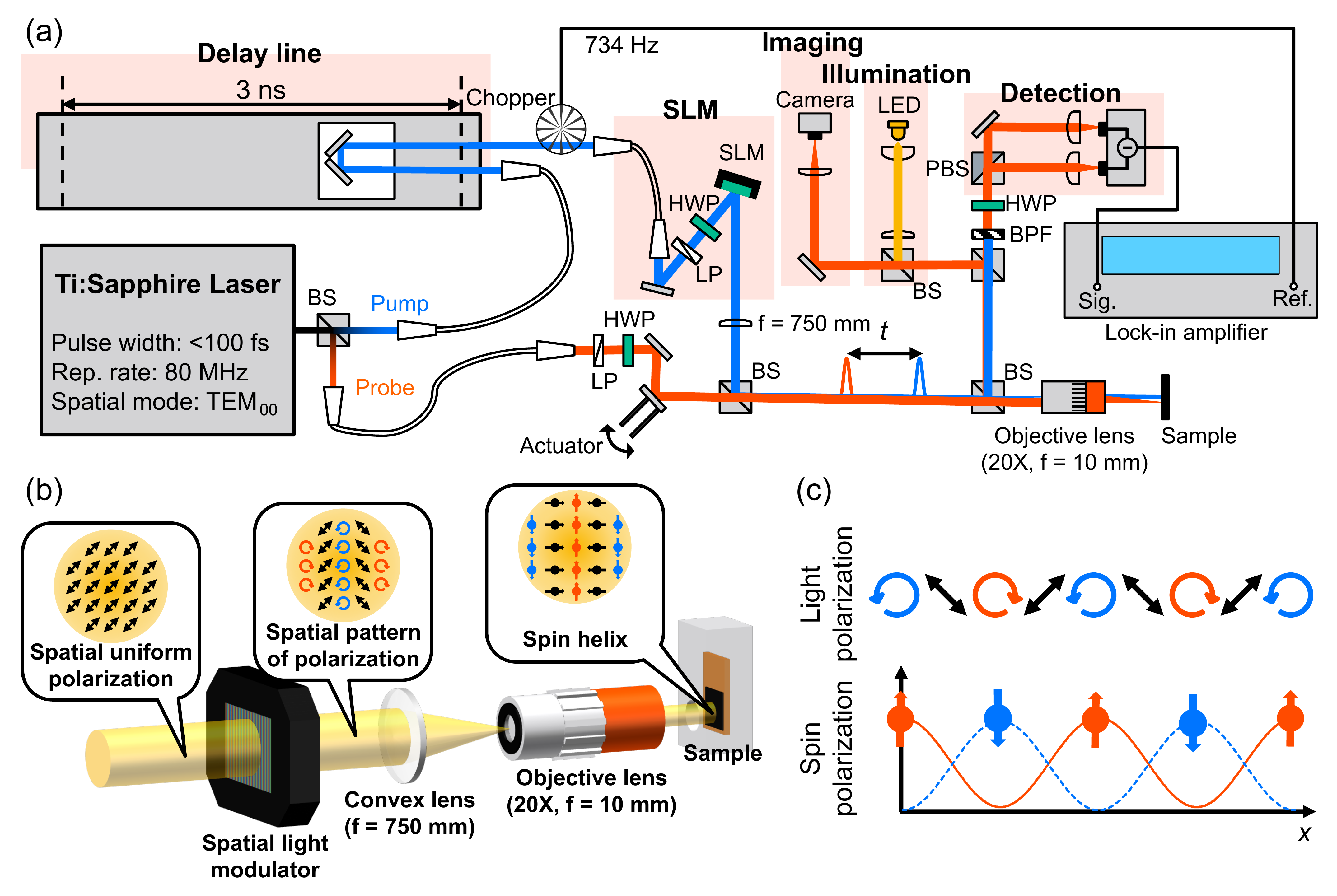}
\caption{(a) Schematic of the optical measurement system. The pump beam is spatially modulated by an SLM, then its beam diameter is reduced by a combination of a plano-convex lens ($f = 750 ~\mathrm{mm}$) and an objective lens ($f = 10 ~\mathrm{mm}$) before being irradiated onto the sample. The probe beam, in contrast, is focused to a small spot on the sample, with spatial scanning performed by an actuator. The reflected light is detected via balanced detection to measure Kerr signals. (BS: beamsplitter, PBS: polarizing beamsplitter, LP: linear polarizer, HWP: half-wave plate, BPF: bandpass filter)\label{fig:fig1a} (b) Detailed schematic of the pump beam. The pump beam initially possesses uniform D-polarization. By spatially modulating this light polarization using an SLM (although it is a transparent arrangement in the diagram, it actually uses a reflective type) and reducing the beam size, a spatial spin polarization pattern (spin helix) corresponding to the spatial light polarization pattern is generated.\label{fig:fig1b} (c) Relationship between the spatial light polarization pattern and the spin helix generated according to the optical selection rule \cite{Cameron1996-xq}.\label{fig:fig1c}}
\end{figure*}

The sample measured was a high-mobility (001) GaAs/AlGaAs QW with a QW width of 20 nm. The structure consisted of the following layers on the GaAs substrate: a 200-nm GaAs buffer layer, a superlattice composed of 60 periods of 18-nm Al$_{0.3}$Ga$_{0.7}$As and 2-nm GaAs, a 100-nm Al$_{0.3}$Ga$_{0.7}$As layer, a 20-nm GaAs QW/35-nm Al$_{0.3}$Ga$_{0.7}$As spacer layer/20-nm Si-doped Al$_{0.3}$Ga$_{0.7}$As doping layer (doping concentration $6\times10^{18}~\mathrm{cm}^{-3}$), and a 5-nm GaAs capping layer. The carrier density was $n_\mathrm{s} = 3.4\times10^{15}~\mathrm{m}^{-2}$. The optical measurement system is illustrated in Fig.~\hyperref[fig:fig1a]{1(a)}. It essentially follows the same configuration as conventional STRKR microscopy, with the addition of a SLM. 
A femtosecond pulsed Ti:Sapphire laser served as the light source. Immediately after emission, the beam was split into two, and each beam’s wavelength was cut out using a diffraction grating and a slit. Two-color pump-probe measurements were conducted with the pump and probe beam center wavelengths set to 799 nm and 810 nm, respectively. 
The pump beam was mechanically delayed in time $t$, and then linearly polarized at 45° (D-polarization) using a linear polarizer and a half-wave plate before entering the SLM. 

The SLM spatially modulates the light polarization, causing the light polarization state to vary spatially in the following sequence: 45° linear polarization (D-polarization) → right-handed circular polarization (R-polarization) → -45° linear polarization (A-polarization) → left-handed circular polarization (L-polarization). This modulation creates a spatial light polarization pattern, which was then focused onto the sample using a combination of a plano-convex lens and an objective lens, leading to the generation of a corresponding spin helix. These configurations and results are depicted in Figs.~\hyperref[fig:fig1b]{1(b)}  and \hyperref[fig:fig1c]{1(c)}. 
The probe beam was aligned to D-polarization using a linear polarizer and a half-wave plate, and then focused onto the sample using the objective lens whereby it was focused to a spot diameter of $\sigma_{\rm{probe}} < 1.3~\si{\micro m}$. Spatial scanning was achieved by adjusting the tilt angle using an actuator and a gimbal mirror. The reflected probe beam from the sample was separated from the pump beam using a bandpass filter centered at 810 nm. The Kerr rotation angle was detected by splitting the reflected beam with a polarizing beamsplitter and performing balanced detection. Additionally, an illumination system and an imaging system were incorporated to verify the focus of the beam on the sample.

\subsection{Control light polarization by SLM}
The SLM used in this experiment was a reflective LCOS-SLM, manufactured by Santec Corporation. An SLM is a device that spatially controls the phase and intensity of light and is composed of liquid crystals and a CMOS chip. The refractive index of the liquid crystals varies with their orientation, affecting the speed of light and thereby altering only the phase component of the light that is parallel to the liquid crystals. The orientation of these liquid crystals can be controlled by a 10-bit (1024 levels) voltage per pixel from the CMOS chip.
Consider the case in which D-polarization is incident on the SLM, assuming perfect light polarization, the phase difference $\delta$, between the components $x$ and $y$, that is, modulated by the SLM, can be expressed as follows using a column vector with $xy$ components:
\begin{eqnarray} \label{eq:1}
\bm{E} = \begin{bmatrix}
E_x \\ E_y
\end{bmatrix} = \begin{bmatrix}
E_0\mathrm{cos}(kz-\omega t) \\ E_0\mathrm{cos}(kz-\omega t + \delta)
\end{bmatrix}.
\end{eqnarray}
If the phase difference $\delta$ is modulated from 0 to 2$\pi$ radians, the light polarization state can transition to R-polarization, A-polarization, and L-polarization, effectively functioning as a wave plate. Given that $\delta$ is proportional to the input voltage $V_\mathrm{in}$, and with the SLM providing a 1024-level gradation, the phase setting resolution is approximately $2\pi/1024~\mathrm{rad} \approx 0.0061~\mathrm{rad}$.

\begin{figure}[htbp]\label{fig:fig2}
\centering
\includegraphics[width=\linewidth]{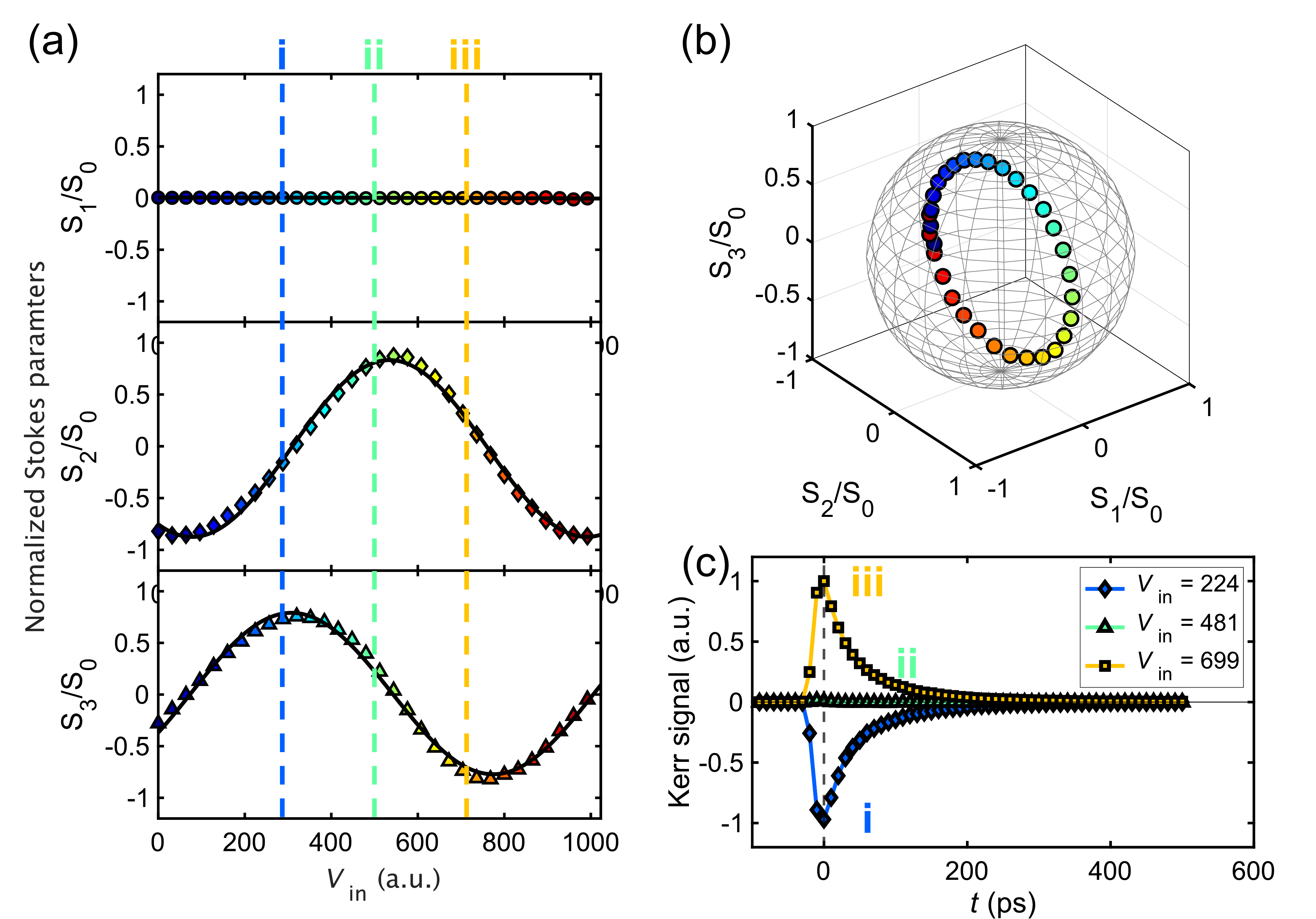}
\caption{(a) Results of measuring the normalized Stokes parameters of the reflected light by uniformly varying the input voltage $V_\mathrm{in}$ across all pixels of the SLM. ($\mathrm{i}$) corresponds to right-handed circular polarization (R-polarization), ($\mathrm{ii}$) corresponds to linear polarization at 45° (A-polarization), and ($\mathrm{iii}$) corresponds to left-handed circular polarization (L-polarization).\label{fig:fig2a} (b) Plot of the data from (a) on the Poincaré sphere.\label{fig:fig2b} (c) Time dynamics of the Kerr signal measured using states of ($\mathrm{i}$) R-polarization, ($\mathrm{ii}$) A-polarization, and ($\mathrm{iii}$) L-polarization. The observation that counter-rotating circular polarizations produce opposite spin polarizations is consistent with the optical selection rule.\label{fig:fig2c}}
\end{figure}

To verify this functionality, D-polarization was incident on the SLM, a uniform input voltage $V_\mathrm{in}$ was applied to all pixels, and the Stokes parameters of the reflected light were measured using a linear polarizer, a quarter-wave plate, and a power meter. 
The first Stokes parameter $S_0$ describes the total intensity of the optical beam; the second parameter $S_1$ describes the preponderance of H-polarization over V-polarization; the third parameter $S_2$ describes the preponderance of D-polarization over A-polarization and, finally, $S_3$ describes the preponderance of R-polarization over L-polarization.
As shown in Fig.~\hyperref[fig:fig2a]{2(a)}, the normalized Stokes parameters ($S_1/S_0$, $S_2/S_0$, $S_3/S_0$) varied periodically with the input voltage $V_\mathrm{in}$, indicating that the light polarization state changed in the following sequence: D-polarization → R-polarization → A-polarization → L-polarization → D-polarization. This confirmed that the phase was modulated linearly with respect to the input voltage $V_\mathrm{in}$ and that the total phase shift exceeded 2$\pi$ radians. Additionally, Fig.~\hyperref[fig:fig2b]{2(b)} shows that the light traced a complete circle on the Poincaré sphere. These results confirm that the SLM functions effectively as a phase plate and can control light polarization.

When TRKR measurements were performed using the generated light polarizations—($\mathrm{i}$) R-polarization($\mathrm{ii}$), A-polarization, and ($\mathrm{iii}$) L-polarization—as shown in Fig.~\hyperref[fig:fig2c]{2(c)}, ($\mathrm{ii}$) A-polarization produced almost no Kerr signal, while ($\mathrm{i}$) R-polarization and ($\mathrm{iii}$) L-polarization yielded Kerr signals with opposite signs. This indicates that the up-spin and down-spin states were optically generated, respectively, confirming that spin polarization was achieved in accordance with the optical transition selection rule.

\subsection{Spatial control of light polarization by SLM}
The SLM used in this study had a panel size of 9.60 mm (H) $\times$ 15.36 mm (V), with a resolution of 1200 px (H) $\times$ 1920 px (V), corresponding to a pixel size of 7.8 \si{\micro m} (H) $\times$ 7.8 \si{\micro m} (V). To effectively utilize the panel, a pump beam with a diameter ($1/e^2$ diameter) of approximately 7.5 mm was input. However, as we aimed to generate spin helices with dimensions on the order of tens of micrometers on the sample, the pump beam was reduced to $\sigma_{\rm{Pump}} \approx 25~\si{\micro m}$, 1/75 of its original size by combining a plano-convex lens with $f = 750 ~\mathrm{mm}$ and a 20X objective lens with $f = 10 ~\mathrm{mm}$. Under this configuration, as shown in Table~\hyperref[tab:tab1]{1}, spatial light polarization pattern with a period of approximately 5 \si{\micro m} is expected by setting the phase to change by 2$\pi$ every 50 pixels on the SLM. In principle, this setup could generate spin helices with periods as small as sub-\si{\micro m}.

\begin{figure}[htbp]\label{fig:fig3}
\centering
\includegraphics[width=0.8\linewidth]{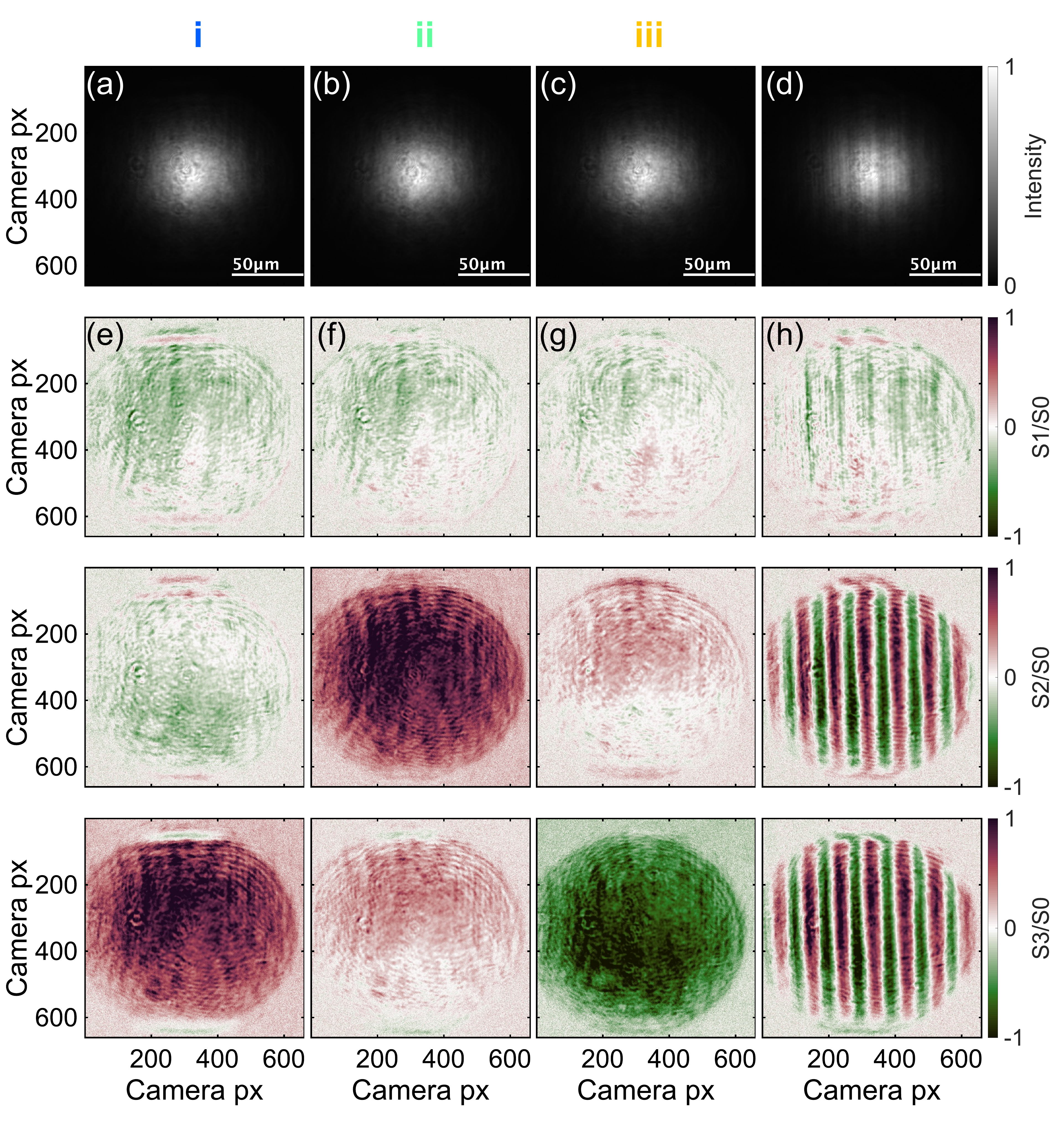}
\caption{(a)-(c) Images of the reflected pump beam from the sample captured with a CMOS camera when input voltages $V_\mathrm{in}$ corresponding to ($\mathrm{i}$) R-polarization, ($\mathrm{ii}$) A-polarization, and ($\mathrm{iii}$) L-polarization were applied.\label{fig:fig3a-c} (d) Image of the reflected pump beam from the sample taken with a CMOS camera when an input voltage $V_\mathrm{in}$ was applied, which caused the phase of the light to shift by $2\pi$ every 96 pixels on the SLM.\label{fig:fig3d}  (e)-(h) Maps of the normalized Stokes parameters corresponding to (a)-(d).\label{fig:fig3e-h} }
\end{figure}

To investigate the spatial modulation of light polarization, the reflected pump beam from the sample was imaged using a CMOS camera using a linear polarizer (analyzer) and a quarter-wave plate. A plano-convex lens with $f = 150 ~\mathrm{mm}$ was positioned in front of the CMOS camera, resulting in a 15X magnification when combined with the 20X objective lens with $f = 10 ~\mathrm{mm}$. This setup is shown in the top row of Fig.~\hyperref[fig:fig3]{3}. Using the same method as before, we calculated the spatial distribution of the normalized Stokes parameters from the light intensity on the CMOS camera for the reflected pump beam. Figs.~\hyperref[fig:fig3a-c]{3(a)-3(c)} display the results for the input voltages $V_\mathrm{in}$ corresponding to ($\mathrm{i}$) R-polarization, ($\mathrm{ii}$) A-polarization, and ($\mathrm{iii}$) L-polarization, while Fig.~\hyperref[fig:fig3d]{3(d)} shows the results for the input voltages $V_\mathrm{in}$ varying in phase by 2$\pi$ every 96 pixels horizontally on the SLM in a periodic manner. The results indicate that the light polarization was modulated in a roughly spatially uniform manner in Fig.~\hyperref[fig:fig3e-h]{3(e)} for ($\mathrm{i}$) R-polarization, in Fig.~\hyperref[fig:fig3e-h]{3(f)} for ($\mathrm{ii}$) A-polarization, and in Fig.~\hyperref[fig:fig3e-h]{3(g)} for ($\mathrm{iii}$) L-polarization. Furthermore, Fig.~\hyperref[fig:fig3e-h]{3(h)} demonstrates that the normalized Stokes parameters $S_2/S_0$ and $S_3/S_0$ vary with a period of approximately 10 \si{\micro m}, and that the phases of each are offset by $\pi/2$. This indicates that spatial light polarization patterns have been generated, and that the period of these patterns is consistent with the spin helix, which is frequently observed in persistent spin helix states.
\begin{table}[htbp]
\begin{tabular}{ccc}
\hline
\multicolumn{2}{c}{On sample} & On SLM     \\ 
Wave number (\si{\micro m^{-1}})   & Wavelength (\si{\micro m})  & Wavelength (px)\\ \hline
0              &              & -          \\
0.1            & 62.83        & 589.1      \\
0.2            & 31.42        & 294.5      \\
0.4            & 15.71        & 147.3      \\
0.6            & 10.47        & 98.17      \\
0.8            & 7.85         & 73.63      \\
1.0            & 6.28         & 58.90      \\
1.2            & 5.24         & 49.09      \\ \hline
\end{tabular}
\label{tab:tab1}
\caption{The relationship between the spatial resolution on the SLM and the wave number and wavelength of the spin helix generated by the spatial light polarization pattern, when the pump beam is reduced in diameter ($1/e^2$ diameter) from approximately 7.5 mm to 1/75 of its original size using a combination of a plano-convex lens with $f = 750 ~\mathrm{mm}$ and a 20X objective lens with $f = 10 ~\mathrm{mm}$.}
\end{table}

\subsection{Generation of a spin helix}
\begin{figure}[htbp]\label{fig:fig４}
\centering
\includegraphics[width=\linewidth]{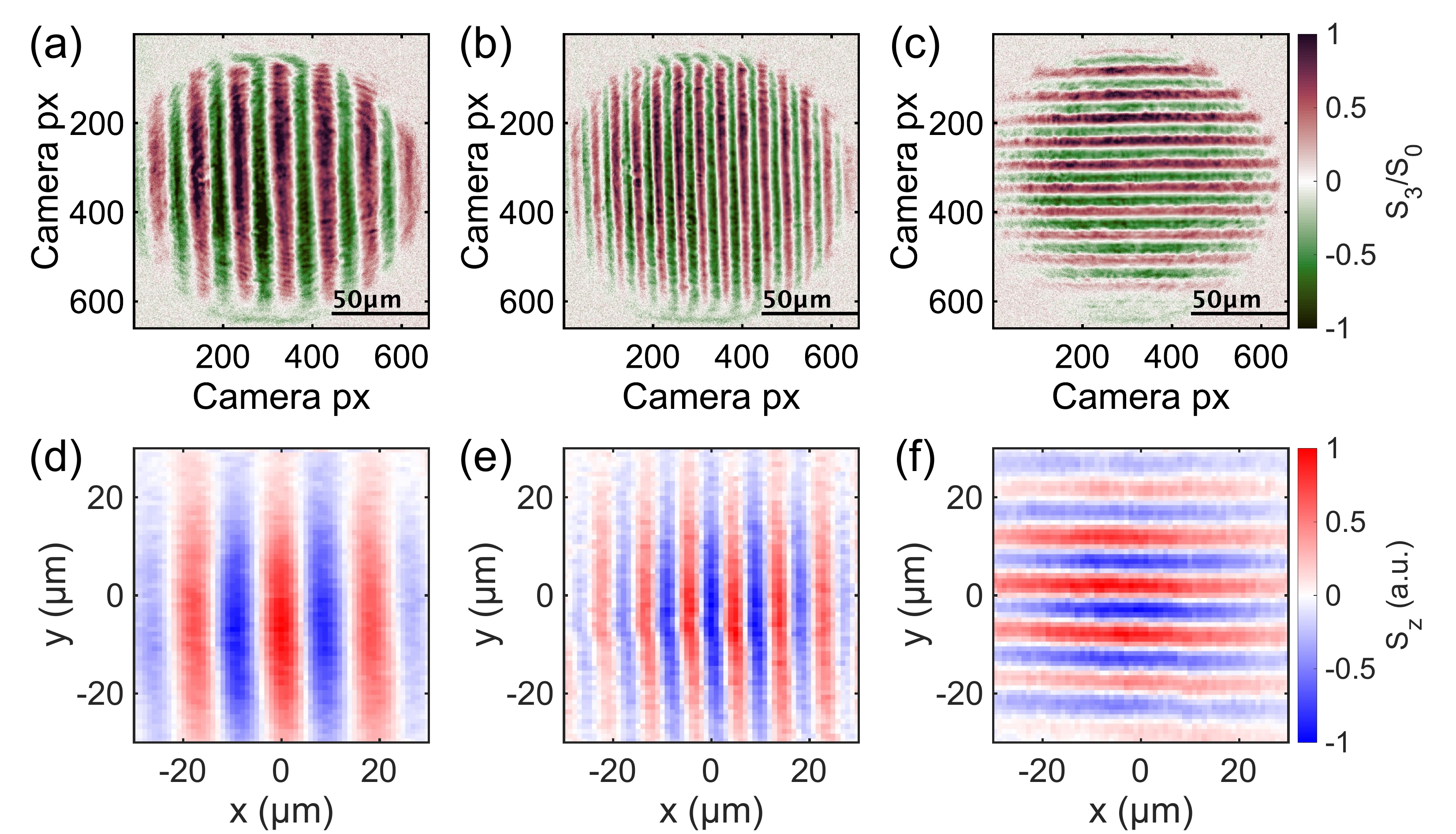}
\caption{(a)-(c) The circularly polarized component $S_3/S_0$ was extracted from the light of the pump beam reflected from the sample when an input voltage was applied to the SLM that induced a phase shift of $2\pi$ every (a) approximately 147 pixels horizontally,\label{fig:fig4a} (b) approximately 74 pixels horizontally,\label{fig:fig4b} and (c) approximately 98 pixels vertically\label{fig:fig4c}. (d)-(f) Spatial maps of the Kerr signal at $t = 0~\rm{ps}$ immediately after excitation corresponding to (a)-(c), respectively.\label{fig:fig4d-f}}
\end{figure}

\begin{figure}[htbp]\label{fig:fig5}
\centering
\includegraphics[width=0.8\linewidth]{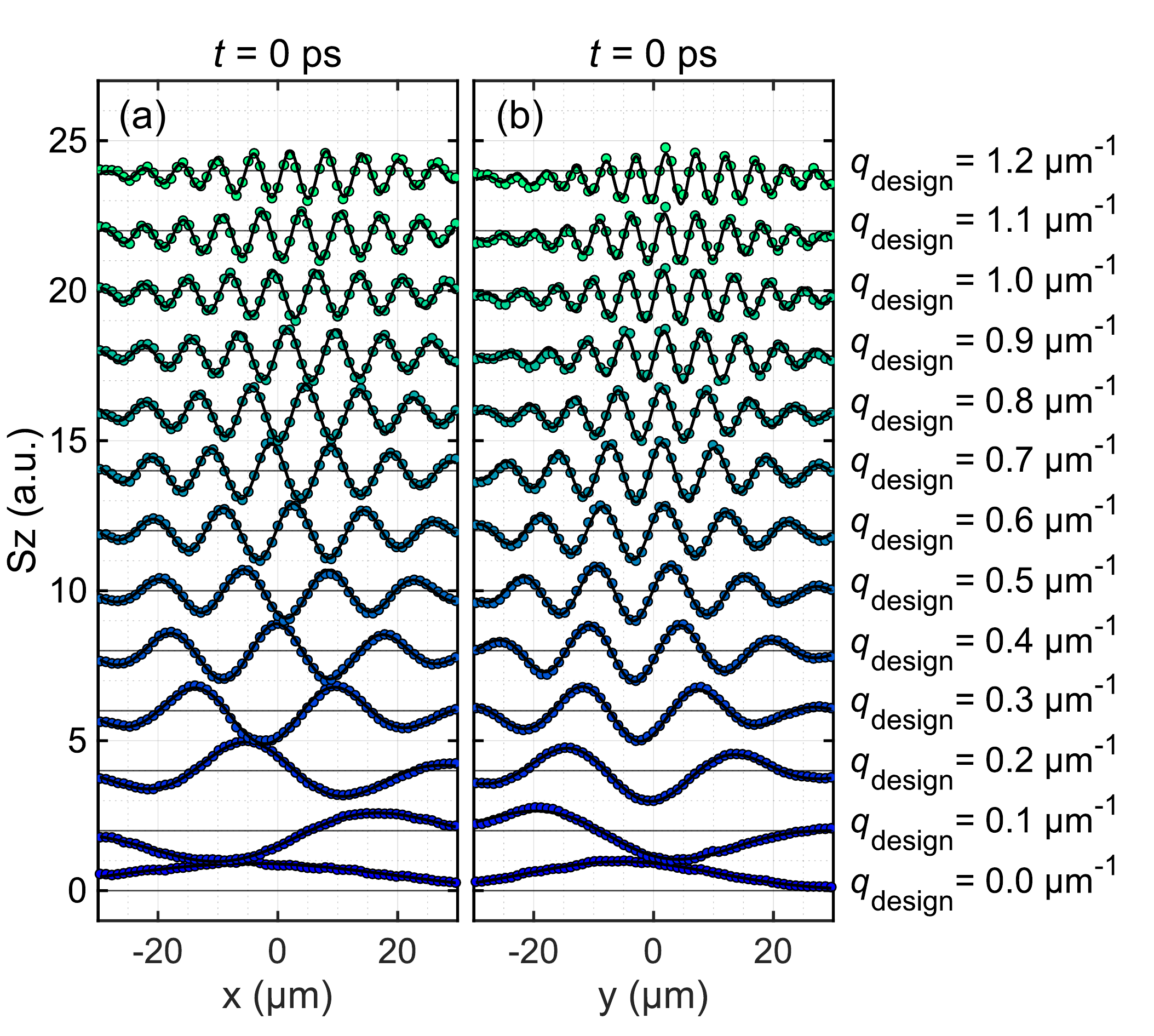}
\caption{(a) Spin helix generated with designed wave number $q_\mathrm{design}$ ranging from 0.0 to 1.2 \si{\micro m^{-1}} along $x // [\bar{1}10]$. (b) Spin helix generated with $q_\mathrm{design}$ ranging from 0.0 to 1.2 \si{\micro m^{-1}} along $y // [110]$.}
\end{figure}

Finally, we measured the spatial map of the Kerr signal immediately after excitation using the spatial light polarization pattern, as shown in Figs.~\hyperref[fig:fig4a]{4(a)-4(c)}, as the pump beam. As a result, we successfully generated spatial spin polarization patterns (spin helices) corresponding to spatial light polarization patterns, as depicted in Figs.~\hyperref[fig:fig4d-f]{4(d)-4(f)}. By adjusting the pixel period controlling the input voltage $V_\mathrm{in}$ on the SLM, we were able to design the wave number $q_\mathrm{design}$ ranging from 0.0 to 1.2 \si{\micro m^{-1}} (or wavelength $\lambda_\mathrm{design}$ from infinity to approximately 5 \si{\micro m}) of the spin helices in both the horizontal and vertical directions, as illustrated in Fig.~\hyperref[fig:fig5]{5}. This result demonstrates that while the wave number $q$ of the spin helix can be modulated by up to a factor of two ($q$ = 0.3 to 0.6 \si{\micro m^{-1}}) through the spin-orbit interaction using a gate voltage \cite{Ishihara2013-eb}, this method allows for the generation over a broader range of wave numbers. Additionally, compared with TSG \cite{Weber2007-py, Koralek2009-mo, Yang2011-rj, Yang2012-er, Yang2012-ue}, the wave number control of this approach is programmable and does not require a phase mask, therefore enabling a more precise and detailed control of the wave number of the spin helices. Although the beam was reduced to 1/75 of its original size for this experiment, it is expected that spin helices with longer or shorter wavelengths could be generated using more pixels and adjusting the magnification through different combinations of a plano-convex lens and an objective lens with varying focal lengths. These findings demonstrate that spatial modulation of light polarization using an SLM is highly effective for generating spin helices.  

\section{ANALYSIS}
\begin{figure}[htbp]\label{fig:fig6}
\centering
\includegraphics[width=\linewidth]{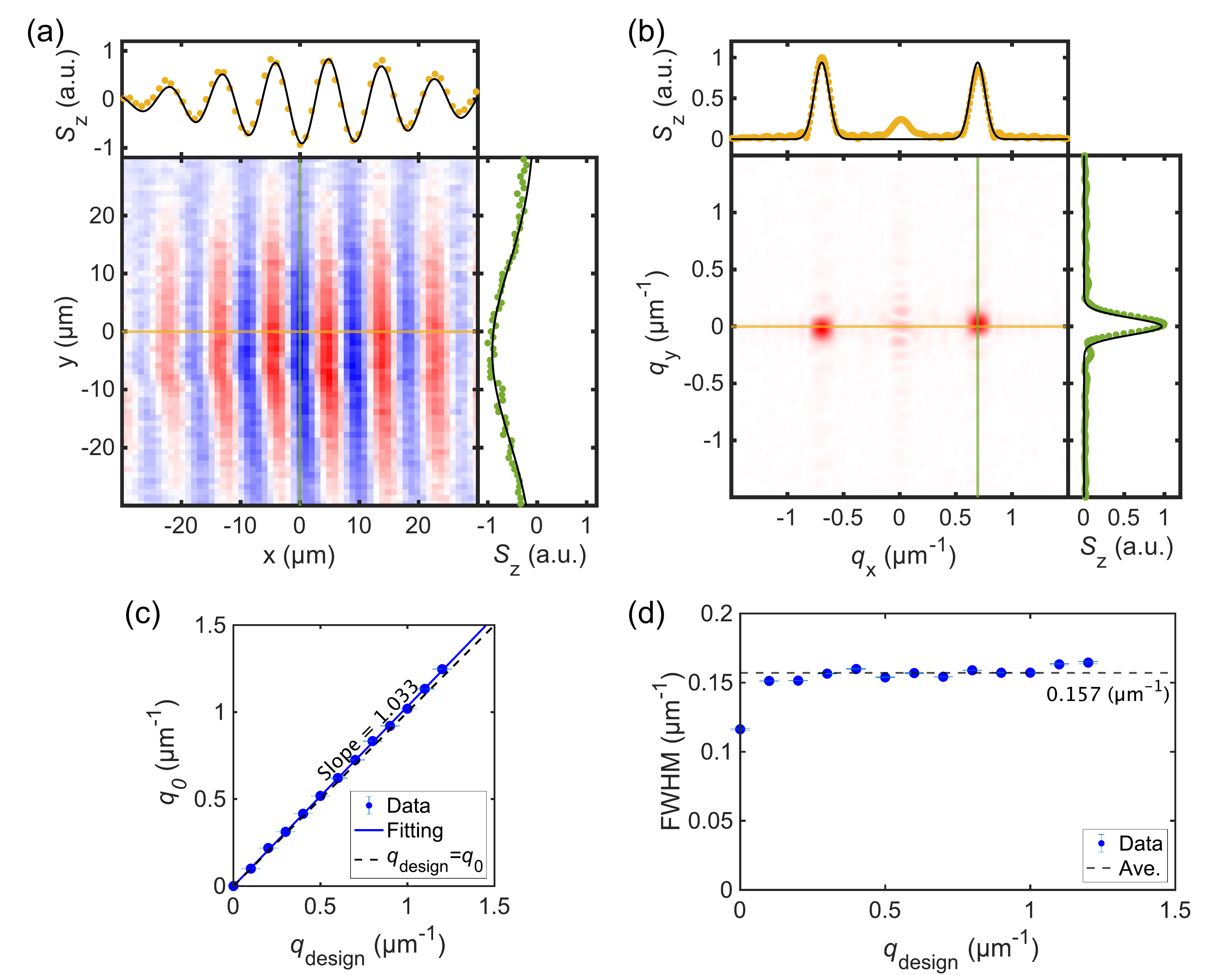}
\caption{(a) Spatial map of the Kerr signal and its corresponding fit when an input voltage $V_\mathrm{in}$ is applied to the SLM, inducing a phase shift of $2\pi$ every 74 pixels in the horizontal direction.\label{fig:fig6a} (b) Wave number space map obtained by performing a two-dimensional Fourier transform on (a) and its corresponding fit.\label{fig:fig6b} (c) Comparison between the central wave number $q_0$ of the generated spin helix, as determined by fitting, and the wave number expected based on the design.\label{fig:fig6c} (d) The full width at half maximum (FWHM) of the generated spin helix in wave number space, as determined by fitting.\label{fig:fig6d}}
\end{figure}

We verified that the generated spin helices matched the design specifications through experimental measurements. The spatial distribution of the excited spin helix, shown in Fig.~\hyperref[fig:fig6a]{6(a)}, consists of a Gaussian envelope combined with a trigonometric wave component, reflecting the original shape of the pump beam. This distribution can be expressed as:
\begin{eqnarray} \label{eq:2}
S_z = A\mathrm{exp}\left(-\frac{\left(r-r_0\right)^2}{2\sigma^2}\right)\mathrm{cos}\left(q_0r+\phi\right),
\end{eqnarray}
where $S_z$ represents the out-of-plane spin component, $A$ is the amplitude, $r_0$ is the center of the Gaussian distribution, $\sigma$ is the standard deviation of the Gaussian distribution, $q_0$ is the central wave number, and $\phi$ is the phase. When this distribution is Fourier transformed in two dimensions, as shown in Fig.~\hyperref[fig:fig6b]{6(b)}, the wave number component can be extracted as:
\begin{eqnarray} \label{eq:3}
\tilde{S_z} = A\mathrm{exp}\left(-\frac{\left(q\pm q_0\right)^2}{2\tilde{\sigma}^2}\right),
\end{eqnarray}
where $\tilde{S_z}$ is the Fourier transform of $S_z$, and $q_0$ and $\tilde{\sigma}$ represent the wave number and its standard deviation in Fourier space, respectively. Figs.~\hyperref[fig:fig6c]{6(c) and 6(d)}. show the fitting results for each wave number. From Fig.~\hyperref[fig:fig6c]{6(c)}, it is evident that the deviation between the generated spin helix wave number and the input wave number value, as determined by the voltage applied to the liquid crystal, is minimal. This confirms that the spin helix was generated with the designed wave number.  Furthermore, the wave number resolution of the generated spin helix, defined as the full width at half maximum (FWHM) for any wave number, is:
\begin{eqnarray} \label{eq:4}
\Delta q = 2\sqrt{\mathrm{ln}(2)\tilde{\sigma}}\approx0.16~\si{\micro m^{-1}},
\end{eqnarray}
which is influenced by the Gaussian shape of the original beam. To generate a spin helix with a higher purity, a beam with a flatter top shape would be required.

\section{CONCLUSION}
In this study, we directly generated spatial spin polarization patterns by controlling structured light. First, we used the SLM to generate various light polarization states, including D-polarization, R-polarization, and L-polarization. We confirmed that these light polarization states varied linearly with the input voltage $V_\mathrm{in}$ applied to the CMOS behind the liquid crystal of the SLM. Additionally, we observed a circle on the Poincaré sphere, demonstrating that the SLM was functioning correctly as a wave plate.
Subsequently, we successfully generated spin helices corresponding to spatial light polarization patterns modulated by the SLM. We also demonstrated that by controlling the input voltage $V_\mathrm{in}$, it is possible to generate spin helices with various wavelengths and wave numbers in both the horizontal and vertical directions, as designed. These results indicate that the spatial modulation of light polarization using an SLM is highly effective in generating spin helices. 

\begin{acknowledgments}
K.K. acknowledges support from the GP-Spin at Tohoku University. (Grant No. JPMJSP2114). This work was supported by the JSPS KAKENHI (Grant Nos. 21H0464 and 21K14528), the JST FOREST, CREST, ASPIRE, PRESTO, and SPRING programs (Grant Nos. JPMJFR203C, JPMJCR22C2, JPMJAP2338, JPMJPR24L4, and JPMJSP2114), and the Cooperative Research Project of RIEC, Tohoku University.
\end{acknowledgments}

\newpage{}
\bibliography{apssamp}

\end{document}